**Title:**

# Automated Materials Discovery Platform Realized: Scanning Probe Microscopy of Combinatorial Libraries

**Authors:**

Yu Liu[1*], Aditya Raghavan[1], Utkarsh Pratiush[1], Maxim Ziatdinov[2], Chih-Yu Lee[3], Rohit Pant[3], Ichiro Takeuchi[3], Pochun Hsieh[4], Albert Suceava[4], Edgar Dimitrov[5], Mauricio Terrones[5], Venkatraman Gopalan[4,5] Ian Mercer[4], R. Jackson Spurling[4], Jon-Paul Maria[4], and Sergei V. Kalinin[1,2*]

[1] Department of Materials Science and Engineering, University of Tennessee, Knoxville, Tennessee, 37996, USA

[2] Physical Sciences Division, Pacific Northwest National Laboratory, Richland, Washington, 99354, USA

[3] Department of Materials Science and Engineering, University of Maryland, College Park, Maryland 20742, USA

[4] Materials Science and Engineering Department, Materials Research Institute, the Pennsylvania State University, University Park, Pennsylvania, 16802, USA

[5] Department of Physics, The Pennsylvania State University, University Park, Pennsylvania 16802, USA

**Abstract:**

Combinatorial materials libraries provide a powerful platform for mapping how physical properties evolve across binary and ternary cross-sections of multicomponent phase diagrams. While synthesis of such libraries has advanced since the 1960s and been accelerated by laboratory automation, their broader utility depends on rapid, quantitative measurements of composition-dependent structures and functionalities. Scanning probe microscopies (SPM), including piezoresponse force microscopy (PFM), offer unique potential for providing these functionally relevant, spatially resolved readouts. Here, we demonstrate a fully automated SPM framework for exploring ferroelectric properties across combinatorial libraries, focusing on binary Sm-doped $BiFeO_3$ (SmBFO) and ternary $Al_{1-x-y}Sc_xB_yN$ (Al,Sc,B)N systems. In SmBFO, automated exploration identifies the known morphotropic phase boundary with enhanced ferroelectric response and reveals a previously unreported double-peak fine structure. In the (Al,Sc,B)N library, ferroelectric behavior emerges at the phase-stability boundary, correlating with variations in morphology and defect concentration. By integrating automated SPM with wavelength-dispersive spectroscopy (WDS) and photoluminescence mapping, we resolve the composition–morphology–defect–property relationships underlying ferroelectric response and demonstrate a pathway toward a multi-tool, high-throughput characterization platform. Finally, we implement Gaussian-process-based single- and multi-objective Bayesian optimization to enable autonomous exploration, highlighting the Pareto front as a powerful framework for balancing competing physical rewards and accelerating data-driven physics discovery.

## 1. Introduction

Combinatorial spread libraries are powerful tools for materials discovery. These libraries typically contain binary or ternary isothermal cross-sections of multicomponent phase diagrams, and more advanced synthesis methods can generate spatially encoded 4D and 5D compositional spaces [1]. This versatility makes them well-suited for both optimizing materials through direct exploration of compositional spaces and advancing physics discovery by exploring property and microstructure evolution [2-10]. Additionally, temperature gradients during synthesis can help reveal the effects of synthesis variables, while localized ion- or laser-based annealing enables broader exploration of the processing and chemical spaces within the selected material systems [8, 11, 12].

The first experiments in combinatorial research date back to the 1960s [13, 14], with renewed interest in the 1990s following the discovery of high-temperature superconductors [3, 4, 11, 15-17]. However, it quickly became apparent that successful combinatorial research requires not only synthesis but also detailed characterization, along with the ability to derive insights from characterization results and use these for subsequent experiment planning or transition towards different fabrication routes. Without the capacity to quantify structural and functional properties of material as a function of composition, progress in materials discovery remains limited [18, 19]. While advances have been made in discovering materials with easily measurable optical properties [20, 21], progress in other material classes has been slower, leading to "combinatorial winters" similar to the "AI winters" in computing. Additionally, although various compositions across a selected compositional space can be generated, these often correspond to uniform synthesis conditions, while optimal synthesis parameters may vary across compositions. Furthermore, questions persist regarding the relationship between the properties observed in combinatorial libraries and those in corresponding bulk phases. Despite these challenges, multiple strategies are emerging to address these limitations, with characterization remaining the primary bottleneck.

Scanning probe microscopy (SPM) methods are exceptionally well-suited to address the characterization challenge. Several classes of SPM techniques are intrinsically quantitative and yield information that directly correlates with functional properties of interest [22-25]. For example, piezoresponse force microscopy (PFM) allows for quantitative measurements of electromechanical responses, directly relating to piezoelectric tensor coefficients and hysteresis loop probing [26-28]. Similarly, light-induced Kelvin probe force microscopy (KPFM) can detect photovoltage, a critical aspect of photovoltaic materials functionality [29, 30]. In many cases, current- and microwave-based SPM techniques also yield quantitative data since conductivity variations in materials often exceed tip-surface contact variability and related topographic cross-talks.

Several other SPM techniques offer composition-dependent proxy signals that are closely connected to core functionalities, even if the underlying mechanisms are not fully understood. For example, electrochemical strain microscopy detects bias-induced strain, providing insights into ion concentration, molar volume, and mobility in electrochemically active materials [31-33]. Additionally, ferroelectric and ferromagnetic domain patterns serve as proxies for phase symmetry and material disorder strength, while even simple topographic measurements can indicate growth mechanisms, phase separation, and other structural characteristics [34, 35].

However, SPM measurements are sequential and can be time intensive. Standard SPM imaging typically requires around 10 minutes per location, although faster SPM techniques are available. Spectroscopy measurements across sufficiently large grids, necessary for statistical averaging, further lengthen the process. This presents a baseline for SPM exploration of combinatorial samples. For large-scale samples and multiple fast measurements, additional factors, such as instrument tuning and tip condition monitoring, can significantly increase experiment duration. Given these constraints, manual grid-based exploration of combinatorial libraries can take 20-30 hours even for binary cases [36], making ternary and more complex compositional spaces prohibitively time-consuming. In addition, the optimization of materials often requires the consideration of competing multiple objectives. For instance, optimizing dielectric constant while minimizing leakage current in semiconductors, balancing critical magnetic field and critical temperature in superconductors, or maximizing piezoresponse amplitude while minimizing coercive field in ferroelectrics. These challenges underscore the need for machine learning (ML) algorithms to assist in controlling the microscope, tuning parameters, and selecting optimal measurement regions based on optical signals [37, 38]. Developing algorithms capable of efficiently exploring compositional spaces and balancing competing objectives, especially beyond traditional grid-based searches, is particularly important.

Previously, we demonstrated the ML algorithms for physics-based exploration of combinatorial libraries [38-41]; however, in that case the exploration is based only on single objective and selection of location was performed by human operators. More recently, we have demonstrated the automated microscopy exploration of polarization switching at single locations and grid exploration of combi libraries [36, 42]. Here, we demonstrate the first fully operationalized, ML-controlled automated experiment in SPM for combinatorial space exploration. We demonstrate the effect of several possible Gaussian process engines on the binary model ferroelectric library Sm-doped $BiFeO_3$ and suggest possible benchmarking strategies for the automated SPM. Finally, we developed a multi-objective Bayesian optimization (MOBO) based automated exploration workflow and tested it on a ternary combinatorial $Al_{1-x-y}B_xSc_yN$ system, showing the power of MOBO in making complicated decisions to balance competing rewards and physics discovery on the Pareto front and revealing the physical mechanisms that lead to the emergence of enhanced ferroelectric properties in this material.

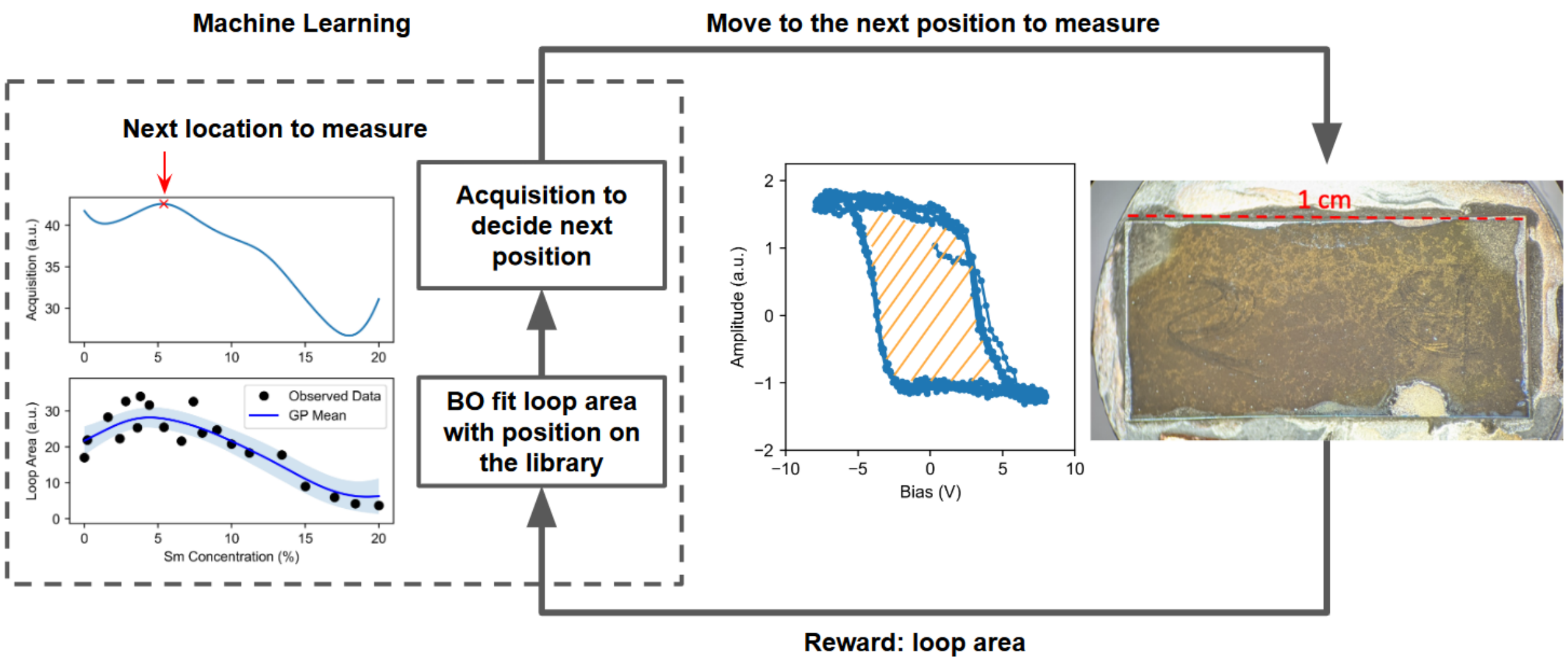

**Fig. 1 | Workflow of autonomous exploration of a binary ferroelectric combinatorial library.** In the autonomous exploration of a ferroelectric combinatorial library, we extract the hysteresis loop areas from the DART spectra as the reward function to represent the ferroelectric property of the library sample. During the exploration process, each iteration starts with fitting the measured loop areas and their locations of measurements with a Gaussian process (GP) model. Then an acquisition function is computed based on the prediction and uncertainty of the GP model. The next location to measure is subsequently determined based on the acquisition function and the experiment policy. The exploration process is iterated until the desired uncertainty level is achieved or the maximum number of exploration steps is reached.

## 2. Results and Discussion

### 2.1. Autonomous exploration of Sm-BFO library

As a first model case, we explored the Sm-doped $BiFeO_3$ (SmBFO) combinatorial library. Here, the BFO side is ferroelectric rhombohedral material, whereas the 20% Sm-doped BFO is orthorhombic non-ferroelectric material. The intermediate concentrations are associated with the formation of the morphotropic phase boundary between crystallographically incompatible materials, often associated with enhanced piezoelectric properties. Previously we used similar samples for a set of the electron and scanning probe microscopy studies. In particular, we used it for human-executed exploration and grid-based exploration via autonomous microscopy [36], thus providing a natural benchmark for this study.

Note that the dependence of the hysteresis loop area on concentration exhibits a non-trivial trend that went unnoticed in previous studies [36]. While we expect the simple maximum of the ferroelectric behavior at the morphotropic phase boundary, here we observe the clear two peaks. We attribute this behavior to the ease of switching of materials close to MPB, that leads to the narrowing of the hysteresis loop due to the switching effect of the Vac probing voltage [43, 44].

These studies illustrate the broad range of information that can be explored over combinatorial spaces.

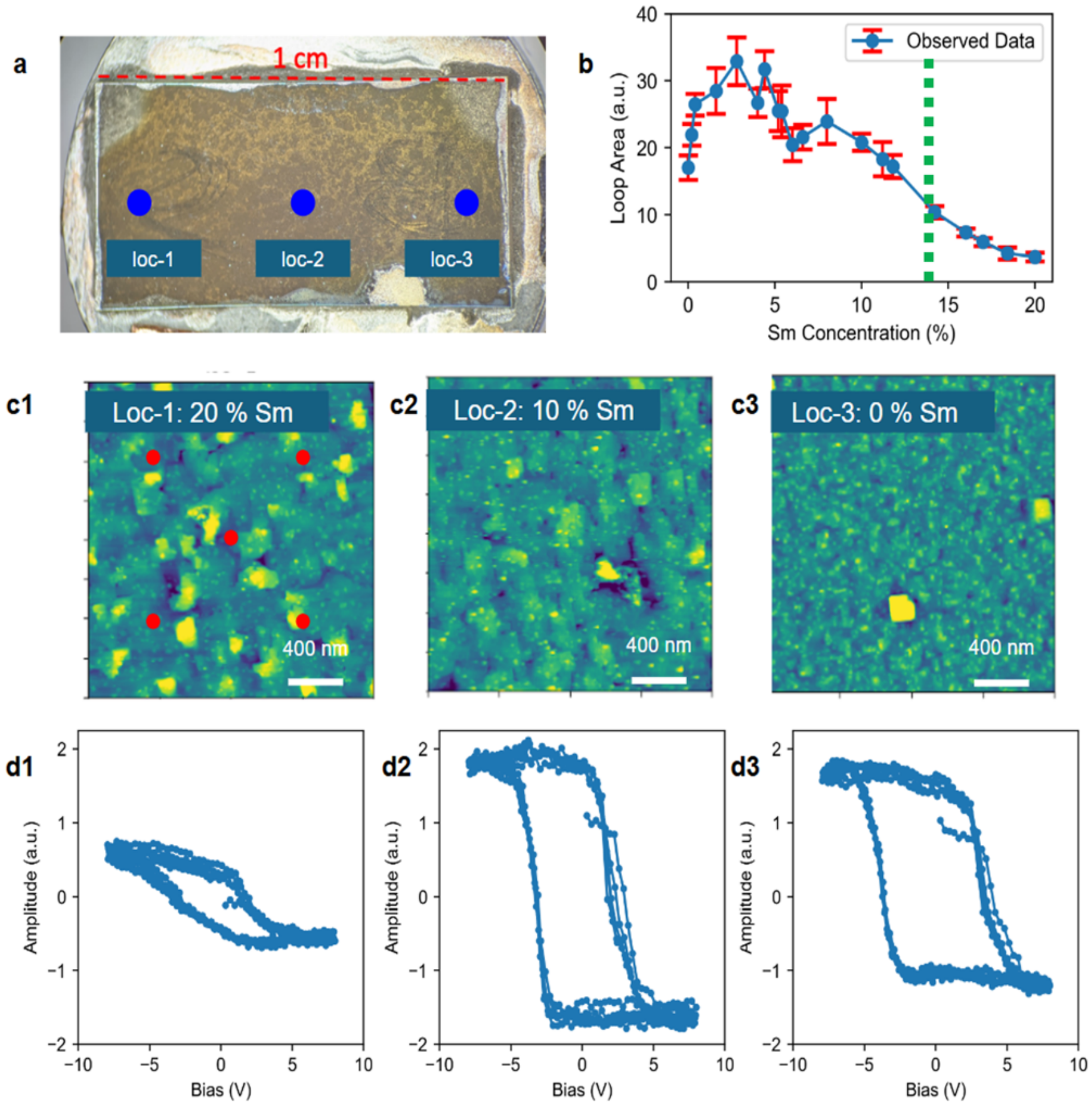

**Fig. 2 | Overview of the combinatorial library $Sm_xBi_{1-x}FeO_3$ (SmBFO). a,** photo of the SmBFO combinatorial library. **b,** grid-search measurements on the SmBFO combi sample. There are 20 locations with 10 DART spectra at each location in the grid. The averaged loop areas are extracted from the DART spectra and plotted as a function of position on the library. The error bar is plotted based on the standard deviation of the loop area measured at each location. The green dashed line indicates the expected composition for the phase separation. **(c1-c3),** topography maps measured at the corresponding locations in **a**. **(d1-d3),** typical piezoresponse amplitude hysteresis loop measured with DART at the same locations in **(c1-c3)**. The red dots in **c1** indicate the spots of DART measurement in the 2 $\mu$m map at each composition.

### 2.2. Implementation of automated exploration workflow

We implemented fully automated exploration of ferroelectric combinatorial library workflow on an Oxford Instrument Asylum Research Jupiter SPM with a fully motorized large-scale sample stage. Because the probe needs to travel large distances above the sample surface, we map out the sample height on a grid and then interpolate to get the estimated sample height across the library before the automated experiment. This sample height information is used to determine the pre-engage height before landing at a new location to avoid tip crashing.

Each new exploration step starts with withdrawing the probe to a constant safe height (here we choose 500 $\mu$m) above the sample surface to avoid tip crashing during sample movement. The next sample position to measure is determined by finding the position corresponding to the maximum in acquisition function based on GP fitting of all previously explored positions. After the sample is navigated to the next position to measure by the motorized sample stage, we first zero the cantilever deflection as the free-air deflection can change at different positions. Then the probe is moved to the pre-engage height of 100 $\mu$m above the expected sample surface according to the interpolated sample height – position relation. After the probe is engaged on the sample surface with a setpoint deflection of 0.5 V, the probe will be positioned to the center of a 2 $\mu$m scan frame to tune the probe in contact, followed by a 2 $\mu$m dual amplitude resonance tracking (DART) PFM [45] map.

To explore ferroelectric properties, 10 DART switching spectroscopy will be performed on a five-point grid in the same area as the DART PFM map, as shown in Fig. 2c1. At each point on the grid, the contact tuning is repeated twice before the spectroscopy to make sure the DART can reliably track the contact resonance despite possible variations of sample topography and surface quality. Each DART spectroscopy consists of 5 full sweeping cycles, and we average the extracted reward functions of the last 4 sweeping cycles to get rid of the large resonance frequency drift effect in the first sweeping cycle. The reward at each step is calculated as the average of the rewards extracted from these 10 spectra with the minimum and maximum rewards excluded as outliers. Similarly, the standard deviation of these 10 rewards is calculated as the reward variation. Based on the rewards and their variations measured at different sample positions, we can either fit a vanilla GP or a measured noise GP to predict the distribution and uncertainty of the rewards across the whole library, and make decisions on where to measure in the next exploration step.

All the operations described above are controlled by Python script through AESPM interface with Jupiter SPM controller to ensure that the whole exploration is fully automated [46]. To make the automated exploration workflow smooth and efficient, it is critical to allocate appropriate wait time for each operation. For operations that require approximately fixed amount of time, like withdrawing the probe, stage movement, probe approaching and tuning, we tested the timing of them repeatedly and used the maximum required duration with a few extra seconds added for safety. We note that in the future the manufacturer provided trigger signal can be used to accelerate these workflows, especially for fast measurement cycles. For the operations that produce data files, such as DART PFM scans and spectroscopy, we kept monitoring the data saving folder. The next operation is triggered once a new measurement data file is saved to the disk.

### 2.3 Compare the performance of vBO and nBO in the automated exploration of SmBFO

In our automated exploration workflow for ferroelectric combinatorial libraries, we aim to identify compositions that maximize loop area while minimizing uncertainty in the loop area-to-composition relationship across the entire library. To achieve this, we define loop area as the area enclosed by the piezoresponse amplitude–bias curves, as shown in Fig. 1. The grid-search of the loop area on SmBFO is shown in Fig. 2b and has been previously reported in [36].

The exploration process starts with five initial seeding measurements, randomly selected on the library. At each seeding position, we measure ten piezo response hysteresis loop spectra using DART PFM, with two measurements at each of five specified spots (red dots in Fig. 2c1). From these ten spectra, we extract ten loop areas, using the average loop area as the scalarizer and the standard deviation as the measured variation of scalarizer.

In automated exploration using vanilla Bayesian optimization (vBO), a GP model is constructed to map scalarizers to their corresponding measurement positions. Subsequently, an acquisition function based on the upper confidence bound (UCB) is computed from the GP mean and uncertainty. The next measurement position is determined by maximizing this UCB acquisition function, followed by moving the probe to the selected position. This iterative process continues with a new GP model and acquisition function trained on all acquired data. The exploration stops once the desired uncertainty level is reached or the maximum number of steps is completed.

In the measured noise Bayesian optimization (nBO) experiment, we continue to use a GP model to fit scalarizers to positions. However, instead of assuming a constant noise distribution, we add an internal GP to capture the variation of the scalarizer. This internal GP provides a noise prediction for each location. Compared to vBO, nBO incorporates experimentally measured variation as model noise, resulting in more precise predictions and improved decisions for subsequent measurement locations.

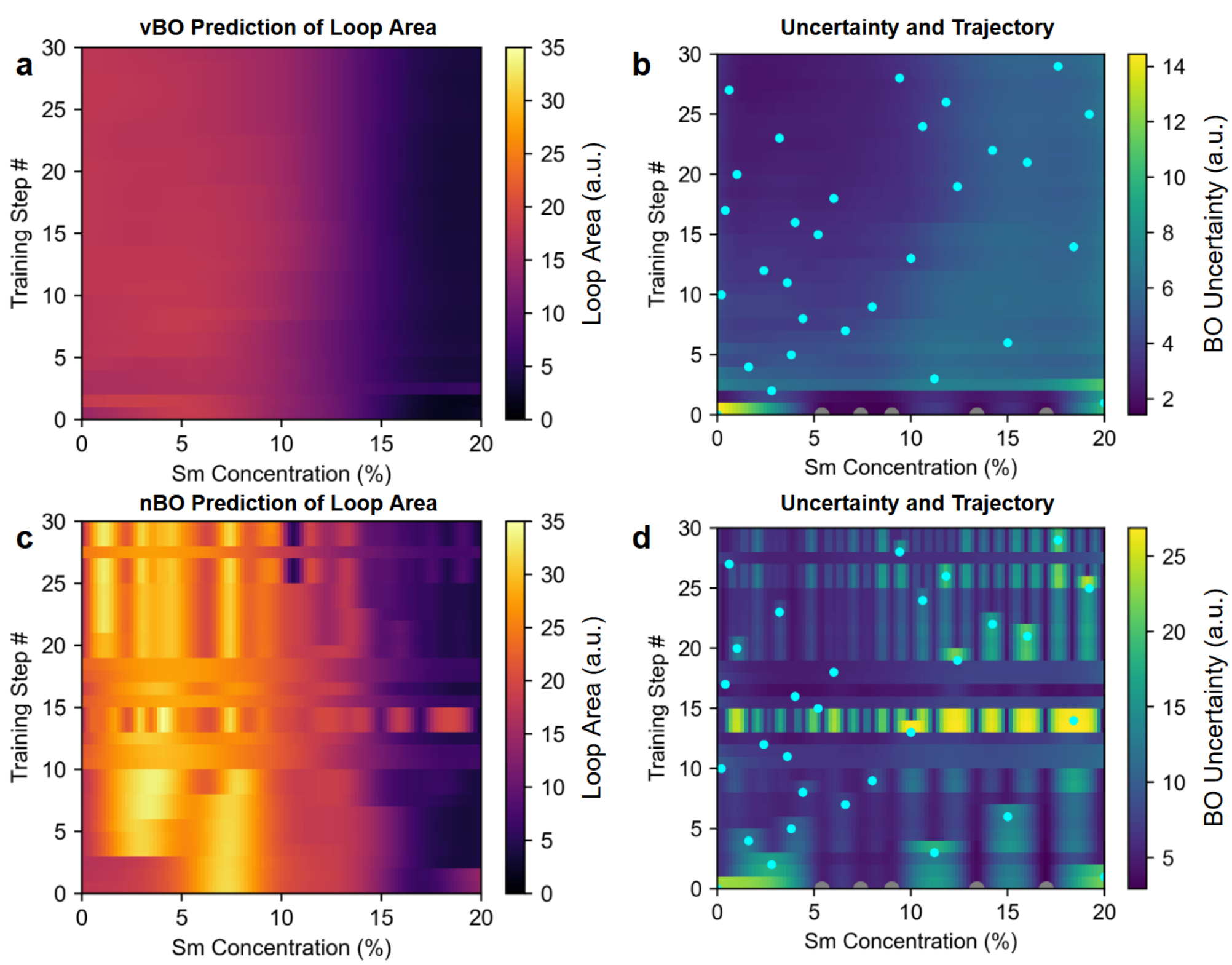

**Fig. 3 | Exploration results of vanilla BO and measured noise BO. a,** predicted loop area as a function of position plotted at each training step for the vanilla BO (vBO). **b,** uncertainty and trajectory of the vBO exploration. The five initial seeding locations are plotted as the gray dots at step = 0. The exploration trajectory is plotted as cyan dots. Model uncertainty is calculated as the difference between the upper and lower bounds of the BO posterior. **c-d,** predicted loop area, uncertainty and trajectory of the measured noise BO experiment. In all the three experiments, there are 5 seeding steps and 30 training steps. To minimize the uncertainty of the loop area – position relation, the acquisition of upper confidence bound (UCB) is used with beta = 100.

In the vanilla Bayesian optimization (vBO) approach, the simple GP model captures the loop area-to-composition relationship with near-ideal accuracy, enabling rapid convergence. With proper seeding, the system's essential features are identified within five steps, as evidenced by the stable predictions of loop areas after step 5 in Fig. 3a. In contrast, the noise-GP model demonstrates significantly slower convergence, characterized by a rapid reduction in kernel length. Here, each local variation in the response is treated as independent of neighboring measurements, leading the algorithm to lose its ability to generalize across locations. This is signaled by the short kernel length and oscillatory behavior at the end of exploration process in Fig. 3c-d. We attribute this outcome to non-material-related factors, including surface topography variations visible in Fig. 1. In this scenario, the noise signal primarily reflects surface inhomogeneity; while the simple GP smooths these variations through its kernel function, the noise GP is more susceptible to instability due to noise spikes. Although these issues can, in principle, be mitigated by adjusting kernel priors or the GP process's mean function, the focus here is on operationalizing and benchmarking specific algorithmic approaches.

In the Supplementary Materials section-III, we have demonstrated the vBO exploration of the hysteresis loop area and ferroelectric turn-on voltage in another binary ferroelectric combinatorial library, $Zn_xMg_{1-x}O$ (ZMO).

### 2.4 Multi-objective automated exploration of ternary combinatorial library (Al,Sc,B)N

We further show the power of our automated SPM through fully automated exploration of a ternary ferroelectric combinatorial library, (Al,Sc,B)N, driven by multi-objective Bayesian optimization (MOBO). These objectives or rewards can be defined as physical properties to directly reflect the goal of experimentation, or as physical descriptors to map out the dependence of the properties on these descriptors [47, 48]. Notably, our MOBO framework is designed to allow users to freely switch between hypervolume-based joint acquisition functions [49, 50], which are excellent for locating the real Pareto front quickly, or separated UCB and EI acquisitions for each reward to get higher freedom and controllability. A detailed explanation of the choices of acquisition functions is included in the Supplementary Materials section-I.

In the MOBO-driven automated exploration of (Al,Sc,B)N library, three rewards are extracted from DART scan and local hysteresis loop: 1), reward 1 is defined as the standard deviation of the topography map to represent the quality of sample growth and phase separation; 2), reward 2 is the averaged piezoresponse amplitude to show the spatially averaged net polarization and polarizability; 3), reward 3 is the loop area measured from point hysteresis loop at the center of the DART scan, serving as a measure of local ferroelectric property. Three individual GPs (Fig. 4) are fitted between the measured rewards and their spatial coordinates respectively, based on which three UCB acquisitions are computed. The next location to measure is determined by maximizing a joint acquisition which is the sum of all the individual UCB acquisition function. In the experiment, there are 20 random seeding locations and 50 active exploration steps.

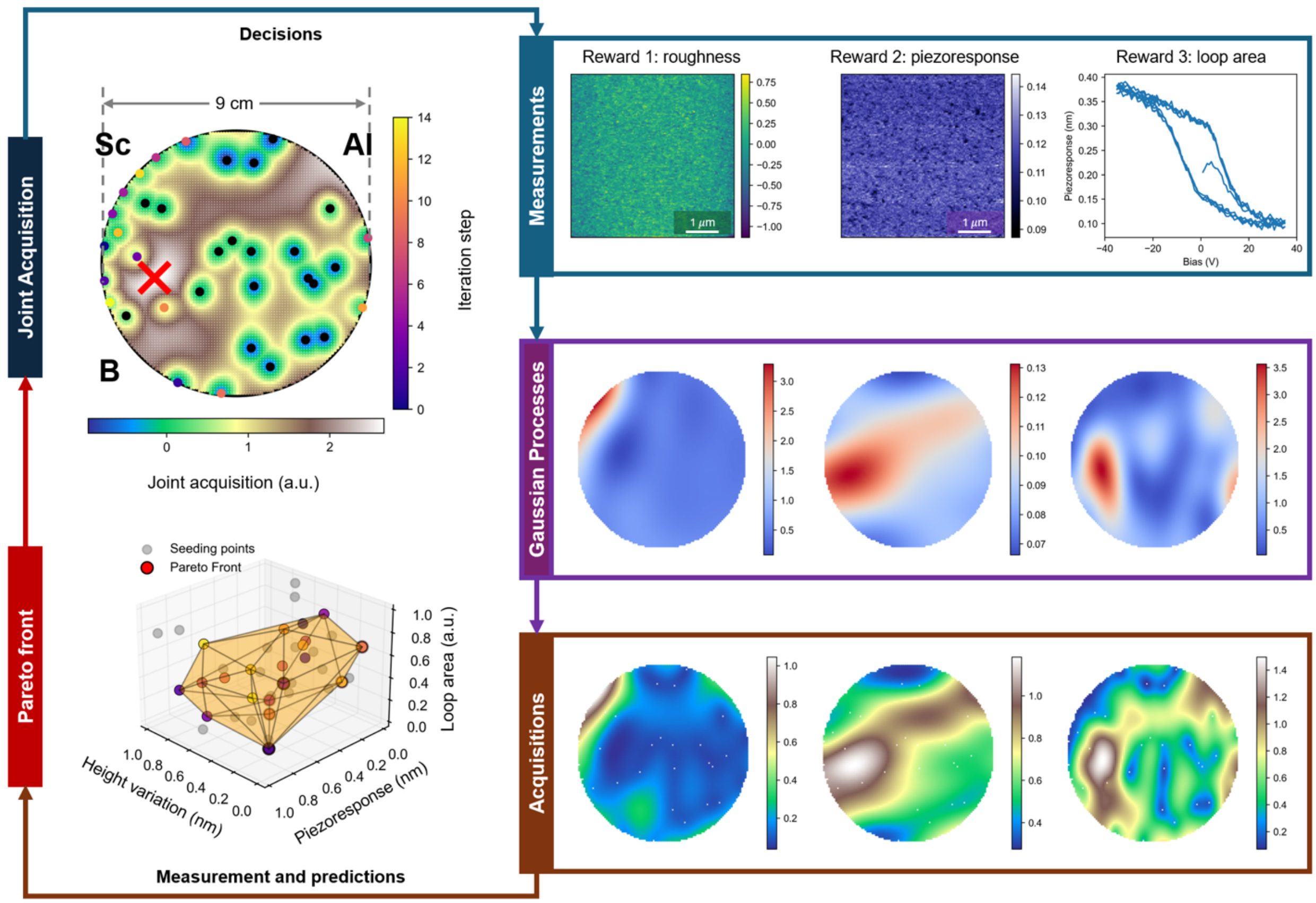

**Fig. 4 | Overview of Multi-objective Bayesian optimization (MOBO) automated exploration of a ternary library.** In a MOBO-driven automated exploration workflow, multiple rewards are extracted from measurements, which are fitted by multiple GPs as functions of locations. A joint acquisition function can be computed directly from all the GPs (hypervolume-based) or based on individual single-reward acquisition functions. The next location to measure is determined by maximizing the joint acquisition function. The Pareto scatter plot reveals the interdependence between each reward.

At the end of the experiment, the resulting scatter plots in Fig. 5a-c show the interdependence between each reward on the Pareto front. The height variation (sample roughness) shows a Pareto front along the 135° diagonal line with respect to piezoresponse (Fig. 5a) and loop area (Fig. 5b), which shows competing behaviors and means a larger sample roughness will lead to smaller ferroelectric response. Such an observation can be understood as sample roughness is a measure of sample growth quality and phase separation. A larger sample roughness corresponds to a larger phase separation and thus leads to reduced ferroelectric response. On the other hand, the loop area shows a cooperative behavior with the averaged piezoresponse along the 45$^{o}$ diagonal line, which indicated that the averaged piezoresponse amplitude generally aligns with the local ferroelectric property measured by hysteresis loop.

In addition to the scatter plot, we also have access to the predicted distribution of each reward across the sample. The predicted distribution of sample roughness (Fig. 5d) shows that when the concentration of Sc element is too high (larger than 25 %), the sample quality degrades quickly, which is likely due to Sc-doping induced phase separation. The comparison between sample roughness (Fig. 5d) and averaged piezoresponse (Fig. 5e) shows that the ferroelectric response is highest along the Al-B line and on the B side and the optimal B doping is larger than 4 % in the (Al,Sc,B)N system, which agrees with previous observations that small amount of B doping can enhance the piezoresponse of AlN system [51, 52]. Interestingly, the largest response in piezoresponse overlaps with the minimum of sample roughness, which suggests that the measured ferroelectric response is potentially determined by the composition through the sample growth quality and phase homogeneity. Besides, the observation that maximum peaks of both the piezoresponse and loop area are pulled towards the Sc source (Fig. 5e-f) indicates that doping Sc into the AlN system can enhance its ferroelectric response. However, it quickly suppresses the ferroelectric properties passing a threshold concentration of 25 % (Fig. 5h), which agrees with previously reported phase separation threshold concentration of 30 % on Si substrate [53] and 10 % to 50 % on AlN and ScN substrates [54, 55]. Finally, the detailed structures in Fig. 5f compared to Fig. 5e shows that the reward of loop area has a more complicated dependence on the composition as it not only correlates to the piezoresponse amplitude but also correlates to other factors like the coercive fields. We further measured the composition of the library by WDS (Fig. S8) and plotted the PFM measured properties as a function of composition in the ternary phase diagrams (Fig. 5g-i). Our automated photoluminescence (PL) characterization (Fig. 5j) in Supplementary Materials section IV further supports that film roughness indicates the piezoresponse through the control of defect concentration, which demonstrates the potential for constructing orchestrated multi-tool autonomous high-throughput materials characterization system where different tools are coordinated by a central AI agent to provide different information about the same material system.

In the attached media file 1, we show the evolution of the MOBO prediction over the active learning steps, which indicates that the model converges well because the predictions barely change after 20 learning steps. The 20 seeding plus 50 learning steps took approximately 6 hours to finish. In comparison, conventional grid-search with the same resolution as the predictions in Fig. 5d-f would require over 1 month to complete for a single library. Such comparison shows that the combination of fully automated SPM and MOBO algorithms can not only provide critical information to downstream applications like sample growth or device fabrication in a high throughput way, but also map out the new physics on the Pareto front effectively.

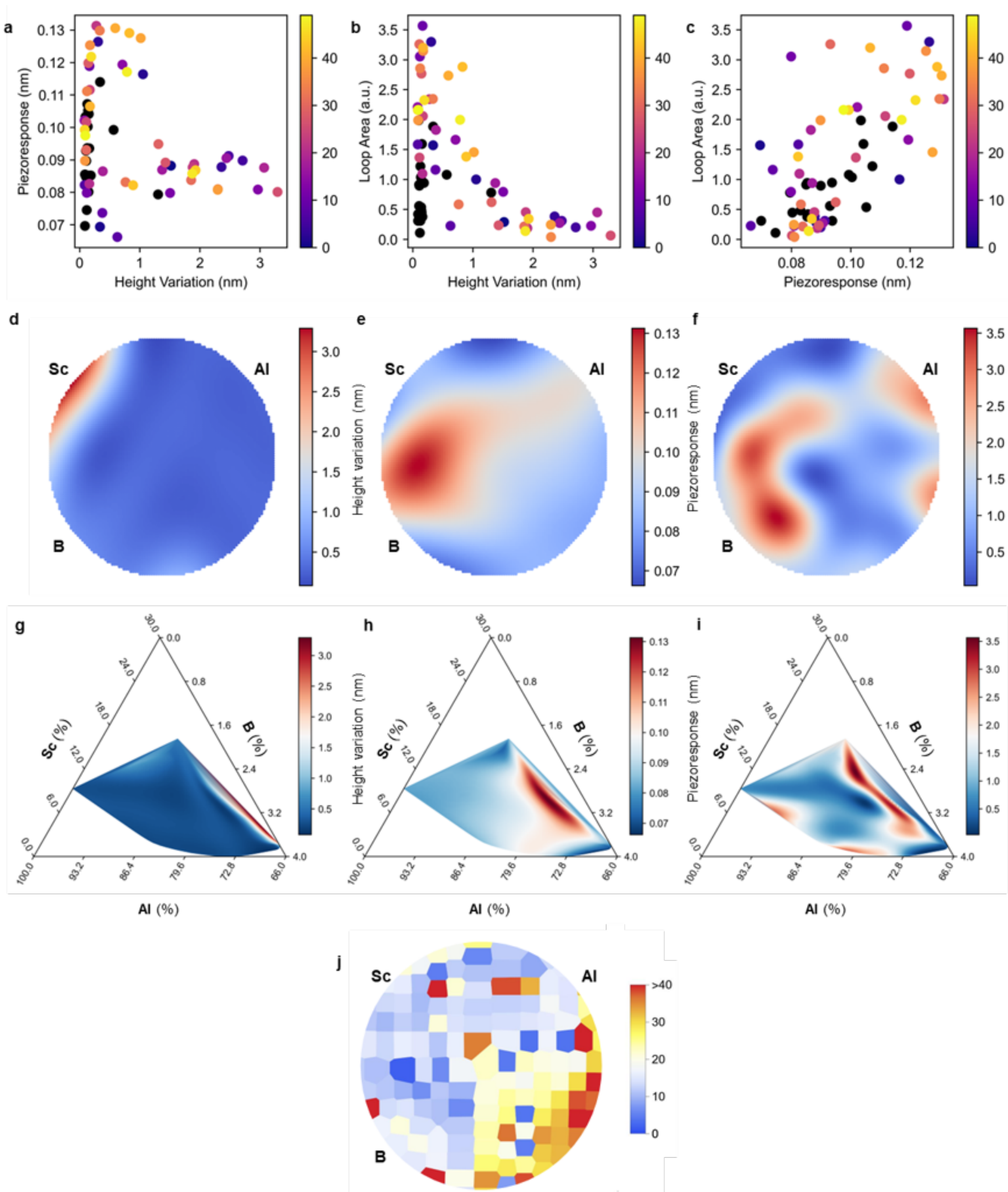

**Fig. 5 | Pareto plot and predicted distribution of rewards. a-c,** Pareto scatter plots for **a** piezoresponse vs. height variation, **b** loop area vs. height variation, and **c** height variation vs. piezoresponse. Here results were obtained through the automated exploration experiment described in Fig. 4. The black dots represent 20 random seeds. **d-f,** predicted distribution of **(d)** height variation, **(e)** averaged piezoresponse, and **(f)** switching spectroscopy loop area across the whole sample. **g-i,** PFM measured properties of **(g)** height variation, **(h)** averaged piezoresponse, and **(i)** switching spectroscopy loop area as a function of library composition in the ternary phase diagram. **j,** Al vacancy concentration map extracted from photoluminescence spectra.

## 3. Conclusion

PFM and, more broadly, SPM offer invaluable capabilities for exploring combinatorial material libraries. By directly detecting material functionalities and leveraging proxy signals, SPM techniques allow detailed characterization across compositional spaces. PFM enables precise measurement of electromechanical properties, making it ideal for studying materials like ferroelectrics and piezoelectrics. These tools bridge the gap in high-resolution, quantitative data acquisition, essential for optimizing material properties in combinatorial research.

Here, we investigated multiple BO algorithms that approach noise handling and scalarization differently, balancing multiple competing rewards with MOBO. The comparison between different single-objective algorithms underscores the impact of noise interpretation and scalarizer choice on the reliability and precision of automated SPM-based explorations. The demonstrated application of MOBO in the exploration of ternary combinatorial system shows the power of MOBO algorithm to map out the interplays between different rewards and their dependence on the composition and quickly find optimal solutions to balance the prediction and uncertainty of multiple competing objectives. Here, the rewards are chosen to represent the growth and characterization parameter, and experimental results suggest that ferroelectric instability emerges at the edge of the phase instability region. It also shows the power of fully autonomous SPM as a high throughput characterization platform for high-dimensional combinatorial libraries, which provides critical information to the material discovery and device fabrication in an unprecedented way.

The methodologies presented here extend beyond PFM, offering potential applications with other types of probes, such as conductive, electrochemical, and magnetic SPM. By adapting BO and MOBO algorithms to different types of signals and scalarizers, this approach can be tailored to analyze various material functionalities. Such flexibility is essential for advancing autonomous materials exploration across diverse combinatorial libraries and expanding the toolkit available for high-throughput characterization of complex materials. By integrating advanced SPM techniques with machine learning algorithms, this framework offers a path toward a faster materials discovery process. Automation reduces the time and labor involved in exploring vast compositional spaces, minimizing resource consumption while accelerating insights into optimal material properties. As this approach continues to develop, it has the potential to transform combinatorial materials research, driving innovation in areas like energy materials, electronic devices, and beyond.

## Methods

The Sm-doped $BiFeO_3$ (SmBFO) thin film used in this paper was grown on $SrTiO_3$ substrate, with 10% Bi replaced by Sm doping. The detailed growth conditions and characterization can be found in previous publications [40, 56, 57].

The ternary (Al,Sc,B)N sputtered film was grown on a 4 inch highly doped n-Si wafer with a resistivity of 0.001-0.005 μΩ*cm acting as the bottom electrode. The n-Si substrate is plasma nitrided with a 1:1 Ar:$N_2$ gas mixture and 50 W stage bias prior to deposition. RF power supplies were used on Al and BN targets while a DC power supply was employed for Sc. The deposition took place in a 24/16 Ar:$N_2$ gas mixture, 350 $^oC$, 1.9 mTorr, and without stage rotation.

Chemical composition of (Al,Sc,B)N thin film library was determined by using wavelength dispersive spectroscopy (WDS) in an electron probe microanalyzer (JXA 8900R Microprobe), with an acceleration voltage of 10 kV and 150 nA. Calibration was done using $Al_2O_3$, $ScPO_4$, $Si_3N_4$ and boron for Al, Sc, N and B respectively with an experimental error margin of <0.3 at. %. The data was interpolated from measurements done every 4mm separation.

The photoluminescence response of the ternary (Al,Sc,B)N sputtered film was mapped using a Photon Systems DUV Raman PL 200 tool with a preset grid. The excitation was 248 nm (5.0 eV) continuous-wave laser with random polarization. Each spectrum was collected using a grating of 300 g/mm to achieve a resolution of 1.2 nm in wavelength with detection performed from 240 nm to 610 nm (2.03 eV to 5.17 eV). A 5x objective was used to achieve a spot size of 120 μm.

The ZMO film used in this study was grown by magnetron co-sputtering at room temperature onto a platinized Si substrate. The Pt layer is 50 nm thick with a 5 nm Ti adhesion layer, both deposited by magnetron sputtering at 300 °C. The ZMO sputtering conditions are 3 mTorr total pressure with gas flows of 18 sccm Ar and 3 sccm 10%$O_3$/90%$O_2$ mix. The ZMO film thickness is 220 nm, grown at a deposition rate of 11 nm/min. The Mg/Zn ratio in the film center is approximately 30/70. Details of ZMO depositions are available in previous publications [58, 59]. Under normal deposition conditions the composition is homogeneous because of a rotating substrate with two sputter guns opposed at 180°. In the present case, the substrate rotation is turned off thus producing a linear variation in Mg:Zn ratio. EDS analysis estimates the composition variation from Mg:Zn=0.198 to Mg:Zn=0.306 across the measurement zone.

For the instrument used in this study, the connections enabling these functions are described in AESPM (automated experiment on SPM) [46]. To make sure readers can reproduce the results shown in this work, we have prepared an open-source Jupyter notebook in this link:

https://github.com/RichardLiuCoding/Publications/blob/main/MOBO_simulation_v2.ipynb

This notebook can be run either online on Colab or locally after downloading, and it contains the core codes to run automated exploration of combinatorial libraries with vanilla BO and measured noise BO.

**Acknowledgement**

We thank Yongtao Liu for helping us take EDS measurement on the ZMO sample. The development of automated exploration workflow of combinatorial library (YL, AR, UP, SVK), growth Sm-BFO and WDS measurements (C-Y L, RP, IT), growth of ZMO and (Al,Sc,B)N samples (RJS, IM and JPM), and PL measurement (PH and AS) were supported by the center for 3D Ferroelectric Microelectronics (3DFeM), an Energy Frontier Research Center funded by the U.S. Department of Energy (DOE), Office of Science, Basic Energy Sciences under Award Number DE-SC0021118. The work at the University of Maryland (growth of SmBFO and WDS characterization of (Al,Sc,B)N) was also supported by ONR MURI N00014172661, NIST cooperative agreement 70NANB17H301, and DTRA CB11400 MAGNETO, Univ. of Maryland.

## Supplement Materials

### I. A tutorial on active learning driven by multi-objective Bayesian optimization

To help the readers understand and apply the MOBO algorithm in their own active learning experiments, we have prepared a Colab Jupyter notebook that readers can play and modify : https://github.com/RichardLiuCoding/Publications/blob/main/MOBO_simulation_v2.ipynb. Basically, users only need to implement the rewards_mobo() and measure() function to adapt to their own instruments and the whole workflow should work automatically.

In the tutorial, we simulated three property maps, superconducting gap size, superconducting in-gap state, and superconducting gap symmetry (Fig. S1a-b). The demonstrated MOBO workflows are designed to find the Pareto front and the distribution of these properties across the map. Here, a Pareto point is a point in the reward (hyperparameter) space such that no solution can be improved in one objective without incurring a cost in another. There are three categories of acquisition functions implemented:

**q-noisy expected hypervolume improvement (qNEHVI) and q-expected hypervolume improvement (qEHVI):** acquisition functions based on expected hypervolume improvement. These are particularly good at finding the real Pareto front quickly. qNEHVI gives better performance compared to qEHVI when the noise is large, as indicated by quick convergence to the Pareto that balanced all the three rewards in Fig. S1d. However, they are not designed to map out the full Pareto front or the detailed distribution of each reward. This can be seen from the poor agreement between model prediction (Fig. S1c) and ground truth rewards (Fig. S1a).

**Upper confidence bound (UCB) and expected improvement (EI):** these options offer flexibility to use different acquisition functions with different hyperparameters for each reward in the tutorial. Basically, the learning policy can be defined individually for each reward, allowing efficient mapping of both the distribution of rewards and the Pareto front (Fig. S1e-f). For example, for property-based reward, an acquisition favoring exploitation is preferred. For physical descriptor-based reward, an explorative acquisition helps discover property-descriptor relationships faster.

**Pure uncertainty-driven:** to get the most accurate predictions on the distribution of each reward, pure uncertainty-driven acquisitions are preferred (Fig. S1g-h).

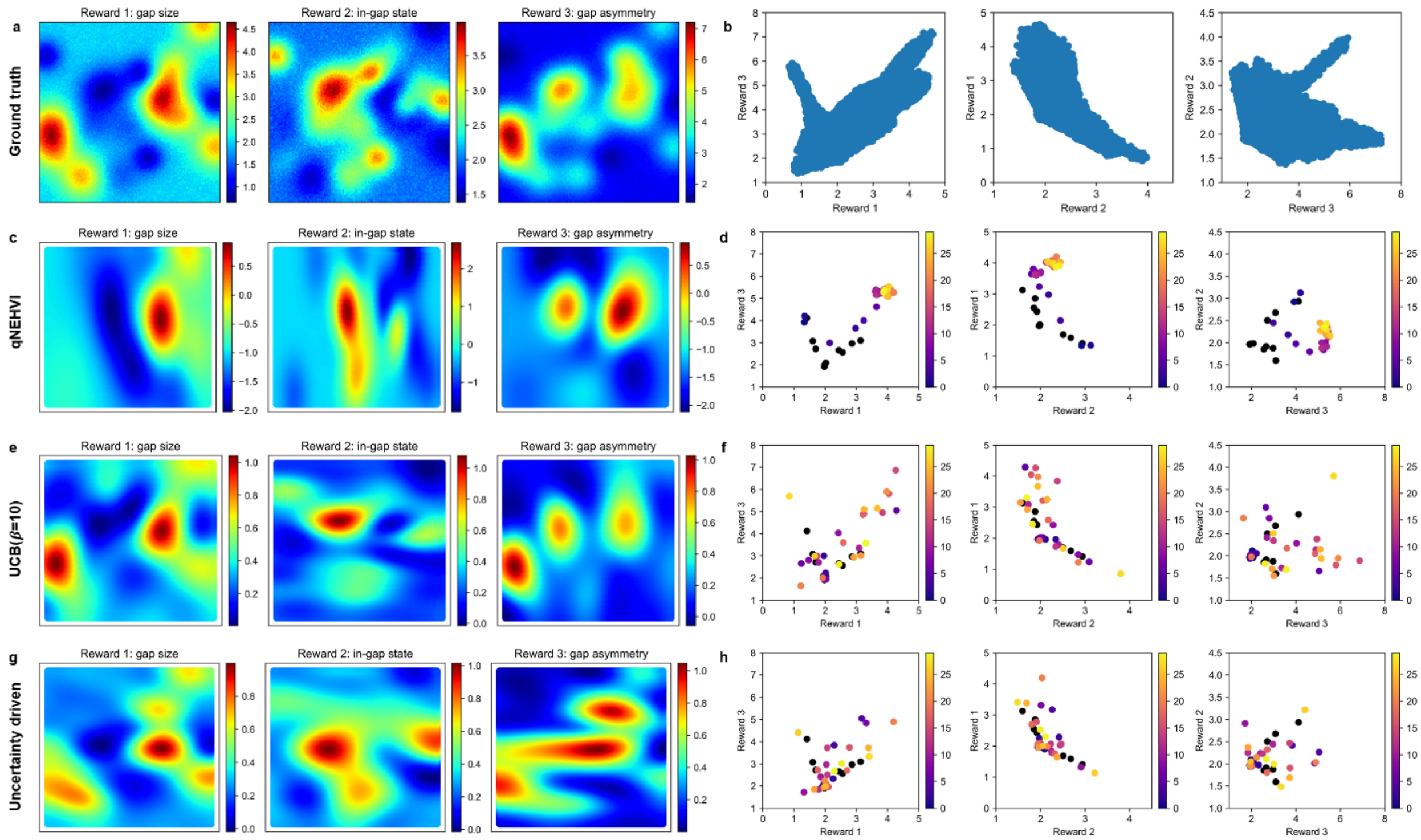

**Fig. S1 | Comparison between different acquisition functions in MOBO simulation. a,** ground-truth simulated maps of superconducting gap-size, in-gap states, and gap symmetry based on local defect types and density. **b,** ground-truth Pareto scatter plots. **c,** predicted distribution of rewards based on active learning simulation with acquisition of qNEHVI. There are 10 random seeds and 30 learning steps. **d,** Pareto scatter plots for the acquisition of qNEHVI. **e-f,** similar results for the acquisition function of UCB with beta=10, and **g-h** with acquisition function of GP uncertainties (pure uncertainty driven exploration). To ensure a fair comparison, the 10 seeding points are kept the same for all the three active learning simulations.

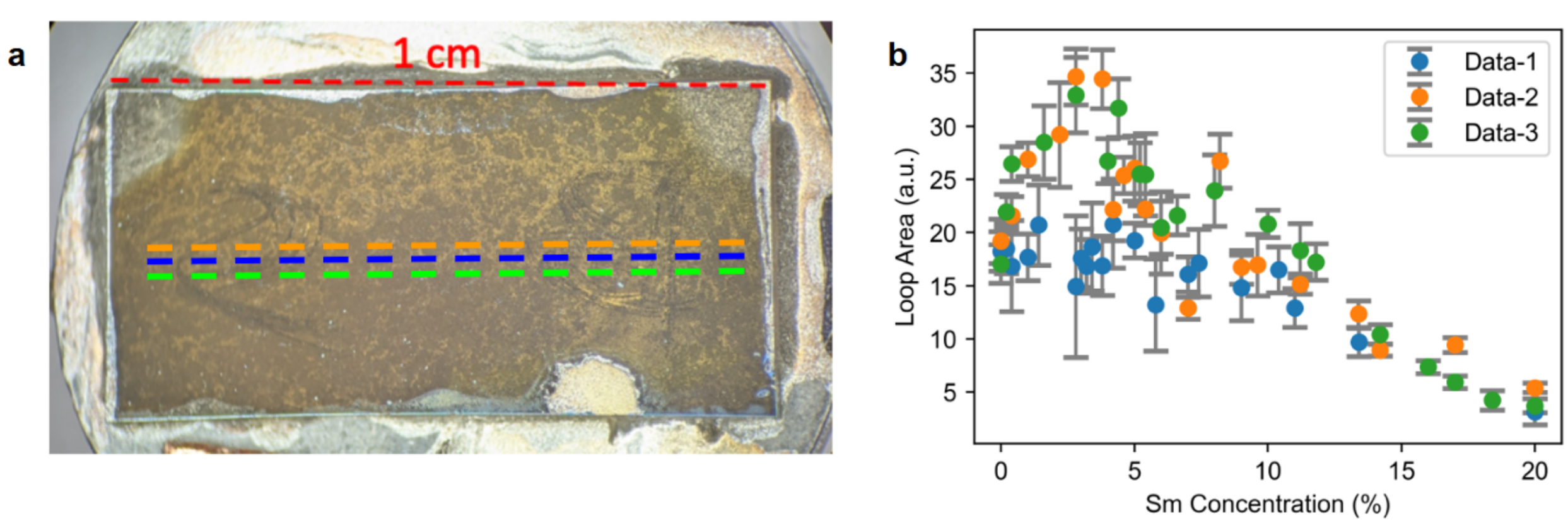

**Fig. S2 | Grid-search of loop area on SmBFO.** a, Picture of SmBFO combinatorial library showing the trajectories of three grid-search DART measurements. Each trajectory is separated by 100 $\mu$m perpendicular to the colored dashed lines in b,

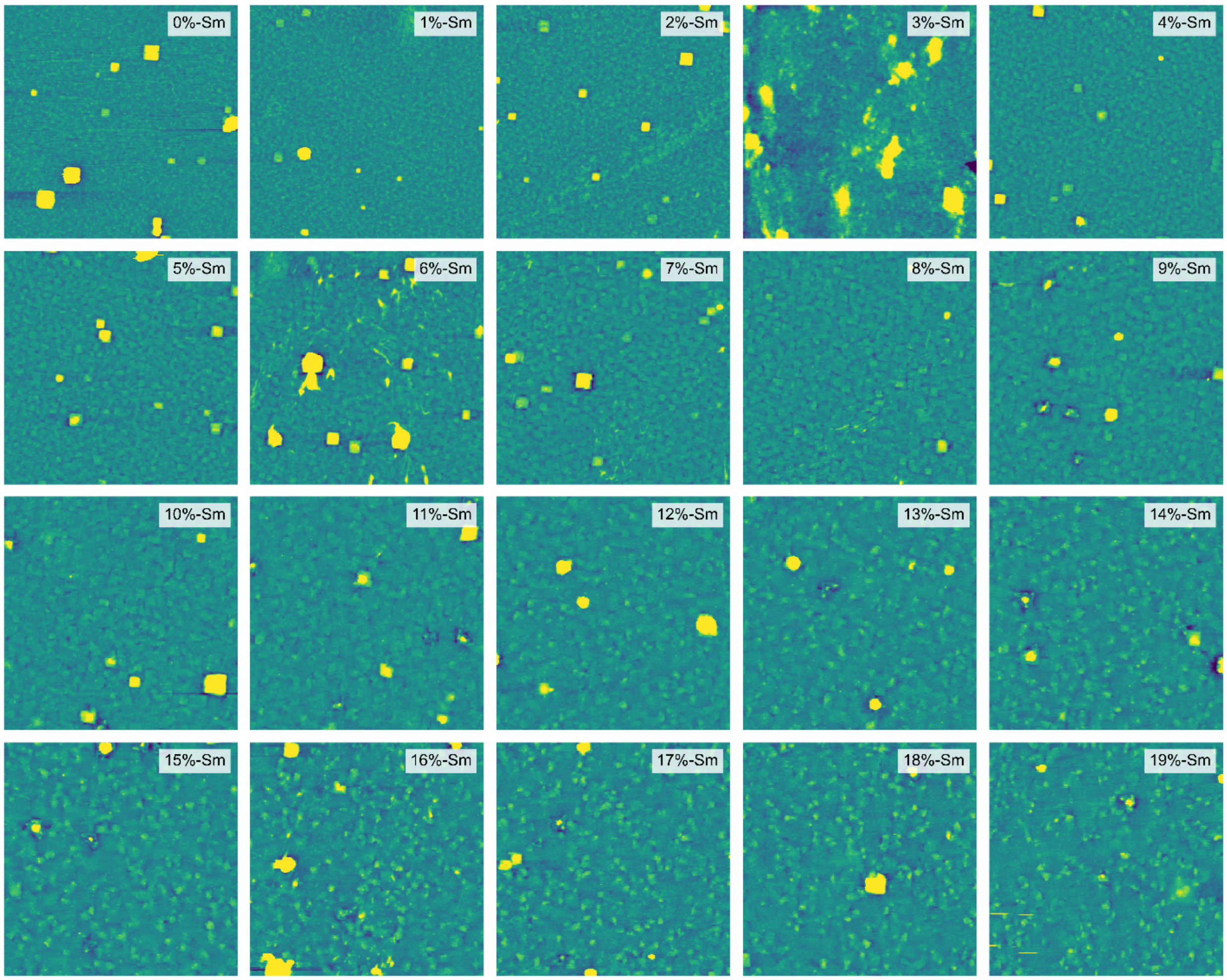

**Fig. S3 | Grid-search of growing quality of SmBFO by topography map.** 5 $\mu$m topography maps acquired on a 20-point grid across the SmBFO combinatorial library. The color scales for all the maps are kept at 26.3 nm for a comparison of sample quality at different compositions.

## II. Comparison between vBO and nBO on a grid

We first compare the performance of vBO and nBO based on grid-search data from the SmBFO library, collected at 20 locations with 10 DART spectra at each location. The 10 DART spectra are distributed over 5 spots within a 2 μm area, as shown in Fig. 2c1, providing a statistical distribution of loop areas. The main characteristics of the average loop area across the composition space shown in Fig. 3a align with expected SmBFO behavior: a gradual increase in loop area from 0 to 7-8% Sm, marking the morphotropic phase boundary (MPB) location, followed by a decline toward the non-ferroelectric composition. To establish a baseline for nBO, we use a Gaussian process (GP) model to fit loop area variations based on position, as seen in Fig. S4b. More grid-search results can be found in Fig. S2.

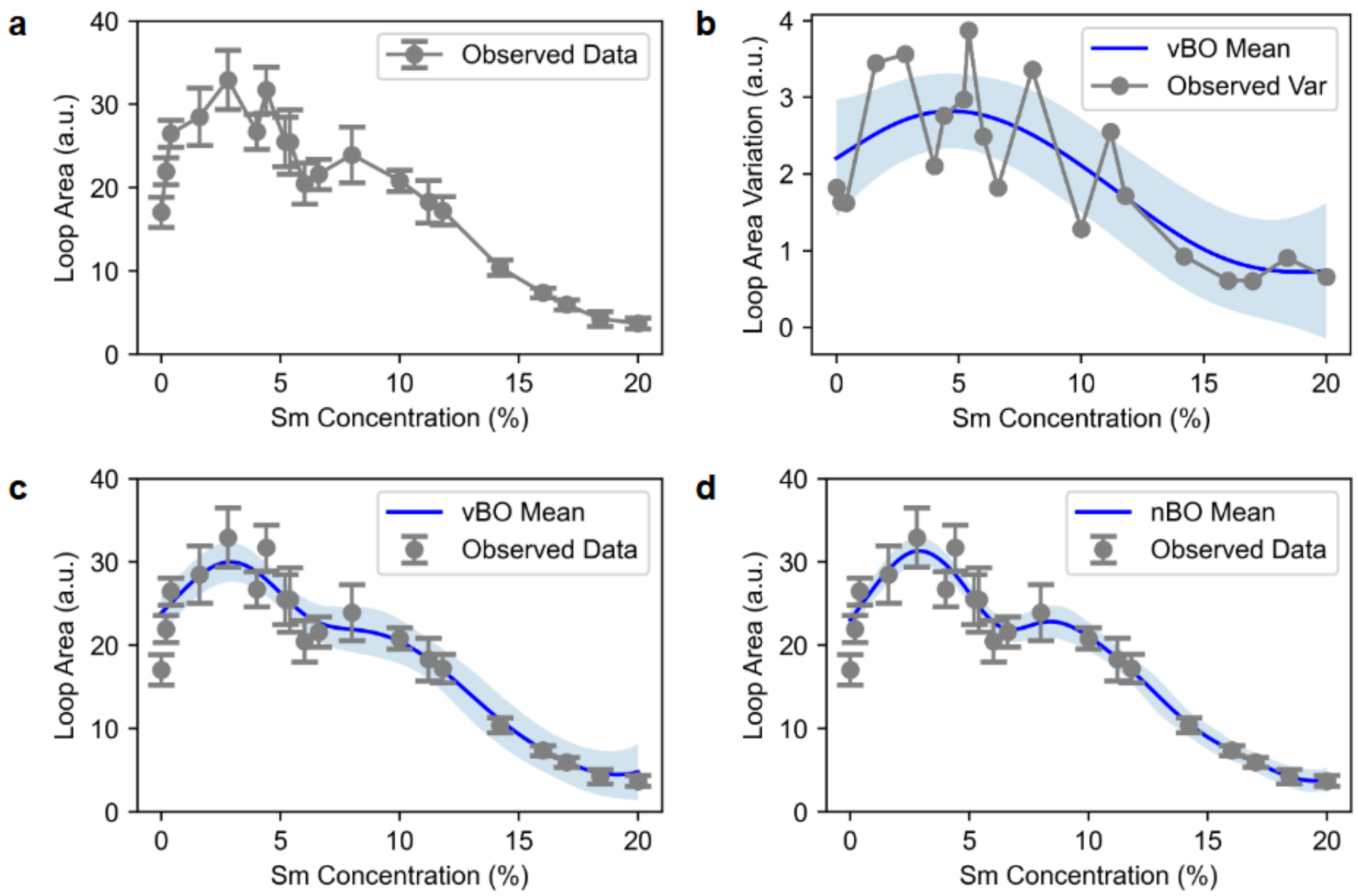

**Fig. S4 | Comparison of vBO and nBO based on the grid-data. a,** exploration of Sm-BFO combinatorial library on a grid. There are 20 locations on the grid. At each location, we measured 10 DART spectra at 5 spots within a 2 $\mu$m area as indicated in Fig. 2c1. The averaged loop areas are plotted as a function of position and the error bar represents the standard deviation of these loop areas at each location. **b,** fitting the measured variation of loop areas as a function of positions by a GP. The confidence interval and the GP mean are plotted on top of the observed variations. **c,** GP fitting of the loop areas. In a vanilla BO, the variation of reward functions is modeled as a distribution function and is inferred only from the variation of reward functions measured at different parameters. **d,** measured noise BO fitting of the loop areas. In addition to using a GP to fit the averaged loop area as **c**, nBO employs an additional internal GP to fit the measured variation (as shown in **b**). Thus, compared to vBO, nBO has a more precise representation of the variation, which helps it make more precise and confident predictions of the reward functions.

In vanilla Bayesian Optimization (vBO), a single GP is used to fit loop areas (Fig. S4c), with uncertainty or confidence intervals derived from the GP model by assuming noise as a constant distribution based on the variability of measured reward functions across locations. In contrast, measured noise Bayesian Optimization (nBO) incorporates an internal GP to fit the experimentally measured variation in reward functions alongside a GP that models the reward functions themselves as shown in Fig. S4b. This approach enables nBO in Fig. S4d to make more detailed and precise predictions, as it captures a richer variation profile compared to vBO. Consequently, nBO exhibits lower uncertainty in predictions than vBO due to its more accurate handling of measurement variability.

However, it is important to note that nBO is more susceptible to measurement outliers because it places greater trust in individual measurements compared to vBO. For instance, if an experimental outlier occurs due to contamination or topographic interference, nBO may make inaccurate predictions based on that outlier. In contrast, vBO can smooth out the impact of outliers, capturing broader trends by maintaining a larger kernel length and allowing for increased noise tolerance. While nBO generally provides more accurate predictions and captures finer details in the data, it is also more vulnerable to occasional outliers.

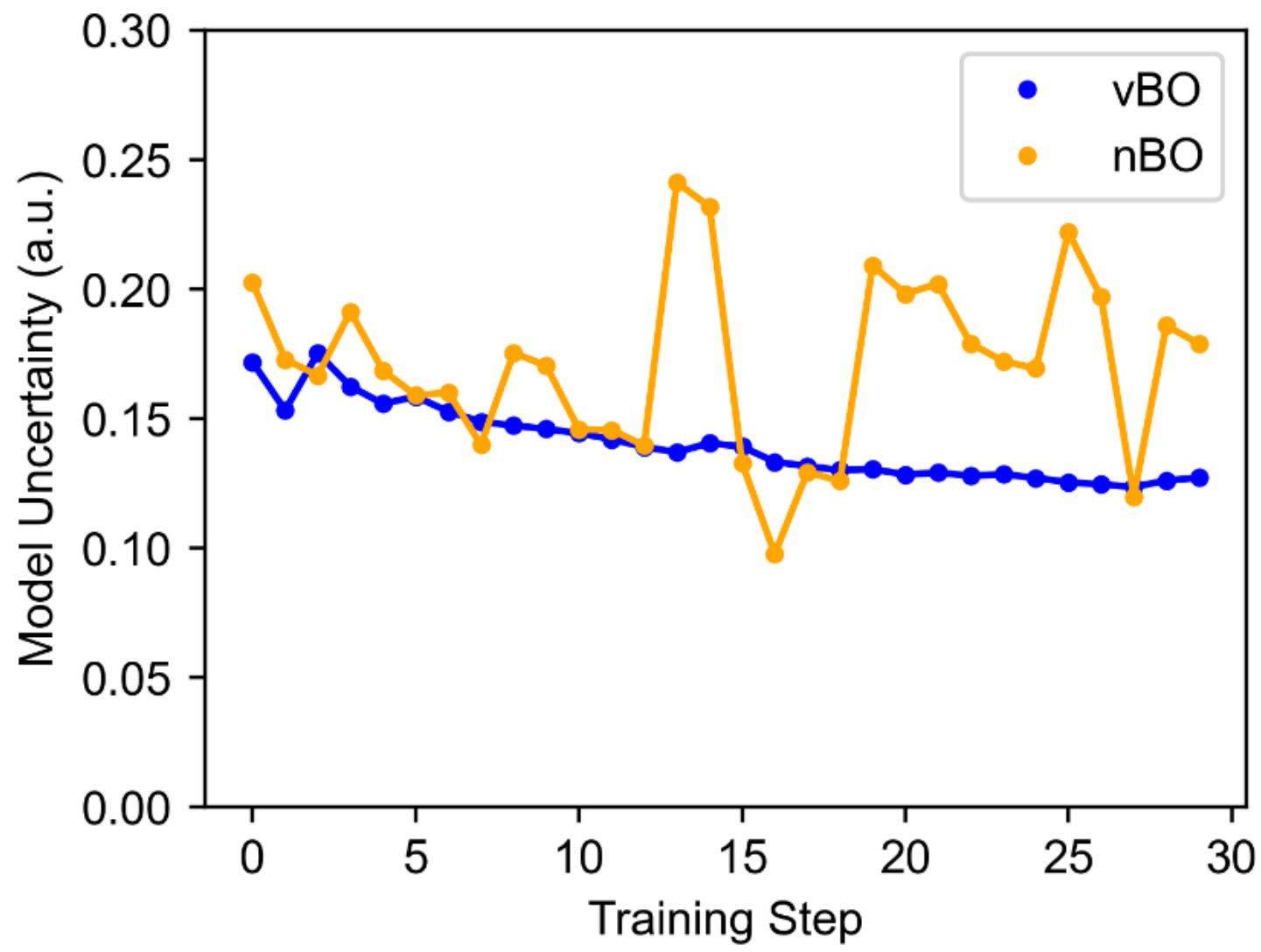

**Fig. S5 | Learning curve for vBO and nBO.** Here the averaged model uncertainty over the whole compositional space at each training step can be used as the stopping criteria.

## III. Autonomous exploration of ZMO library

After we tested our auto-exploration workflow on SmBFO, we further employed our workflow to explore a ferroelectric combinatorial library $Zn_xMg_{1-x}O$ (ZMO), a recently discovered wurtzite ferroelectric material [58]. Here instead of exploring the loop area as a function of composition, we extracted the loop height, defined as the amplitude difference at zero bias in the piezoresponse amplitude–bias curves (the red arrow in Fig. S6b), as the reward. As shown in the grid-search of loop height in Fig. 5c, the measured loop height shows little difference across the library.

In the automated exploration workflow, we used 5 seeding steps plus 20 exploration steps. At each step, 10 DART spectra were measured at 5 grid points in a 2 $\mu$m area. Because ZMO has a higher coercive field compared to SmBFO, we swept the sample bias between +/- 60 V here. As shown in Fig. S6d, although there were a few outliers in the beginning of the exploration, vBO was able to predict a loop height – composition relation well aligned with the grid-search results.

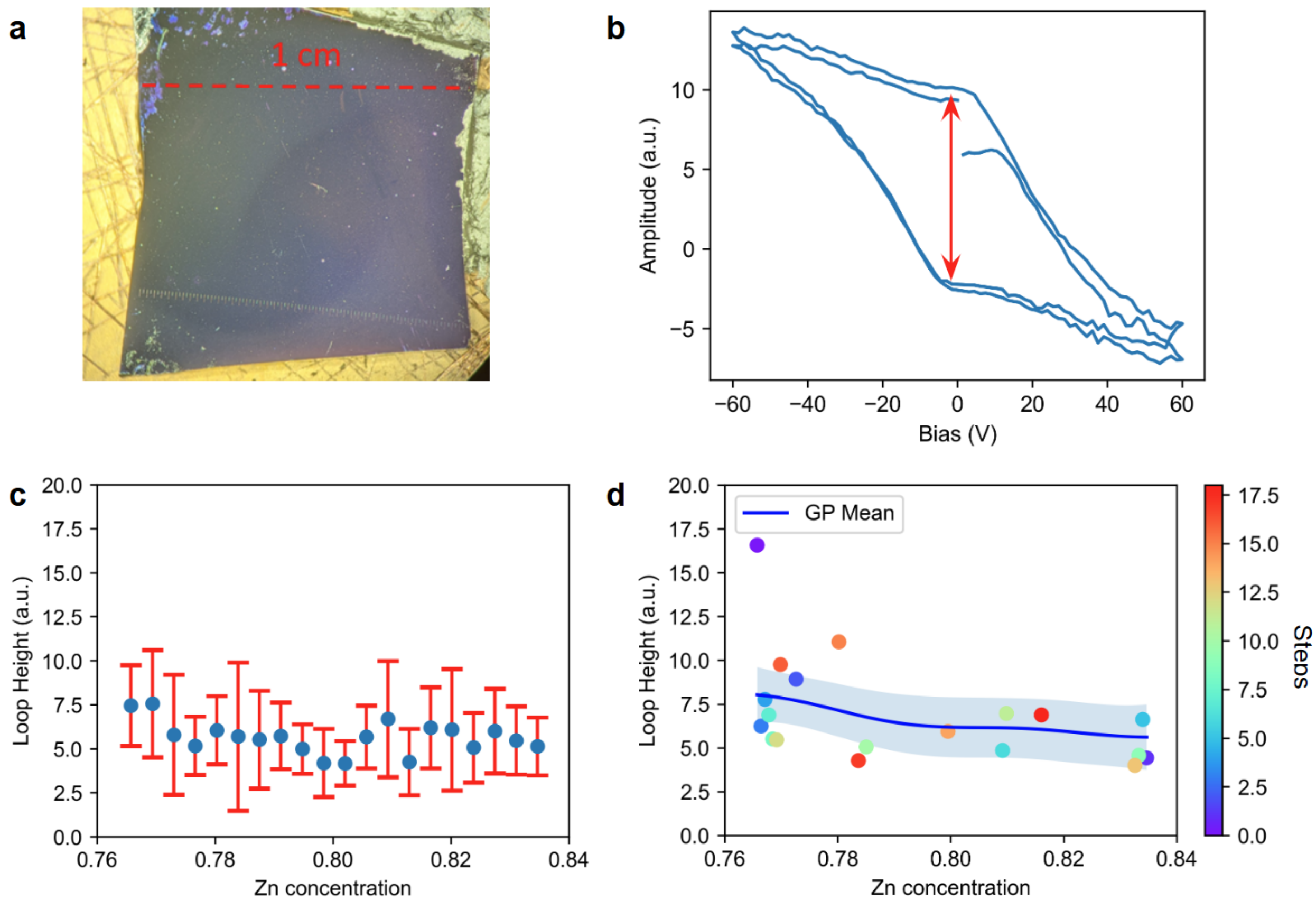

**Fig. S6 | Automated exploration of ferroelectric loop height in $Zn_xMg_{1-x}O$ (ZMO) combinatorial library. a,** picture of the ZMO library. The composition variation is parallel to the red dashed line. **b,** typical piezoresponse amplitude vs. bias curve measured by DART. In this experiment, we define the loop height as the amplitude difference at zero bias and use it as the scalarizer for the auto-exploration. **c,** grid-search of the loop height at 20 positions across the library. **d,** auto-exploration of ZMO library with vBO. Here we used 5 seeding points and 20 exploration steps.

As another example, we explored the turn-on bias in the ZMO library with our automated exploration experiment. As shown in Fig. S7a-b, ZMO exhibits different loop sizes and shapes at different sweeping bias. Between 40 V and 50 V sweeping bias, there is a qualitative change of the phase loop as there starts showing 180-degree full phase flip. This loop opening in the phase is also accompanied by an opening in the amplitude curve in Fig. S7a.

To locate the turn-on voltage precisely, we designed a binary search experiment: at each exploration step, we performed 6 DART spectra. As shown in Fig. S7c, we started with an initial sweeping bias of 40 V and an initial voltage step size of 20 V. At the *i*-th step, the voltage step size is reduced to $20/2^{i-1}$ V, and the sign of the voltage step size is determined by whether there is a 180-degree full phase flip observed at the previous step. If we observed a 180-degree phase flip, it means that we already exceeded the turn-on voltage, so we need to reduce the sweeping bias for the next search (a "—" sign). If there is no 180-degree phase flip observed, then we will increase the sweeping bias for the next search. After 6 searching steps, the turn-on voltage can be found with a precision of 0.625 V.

In this experiment, we used vBO as the ML agent. We started with 5 seeding points and 25 exploration steps. At each step, we performed 6-step binary search at 5 locations on a grid in 2 $\mu$m area, similar to Fig. 2c1. The averaged turn-on voltage of these 5 binary search experiments is used as the reward function. We excluded the rewards greater than 75 V from the vBO fitting as this indicates that either the sample is not switchable or it requires too-high voltage to switch at that location, which is dangerous to the probe.

As shown in Fig. 7d, the automated exploration experiment gives consistent turn-on voltage between 50 V and 60 V across the library.

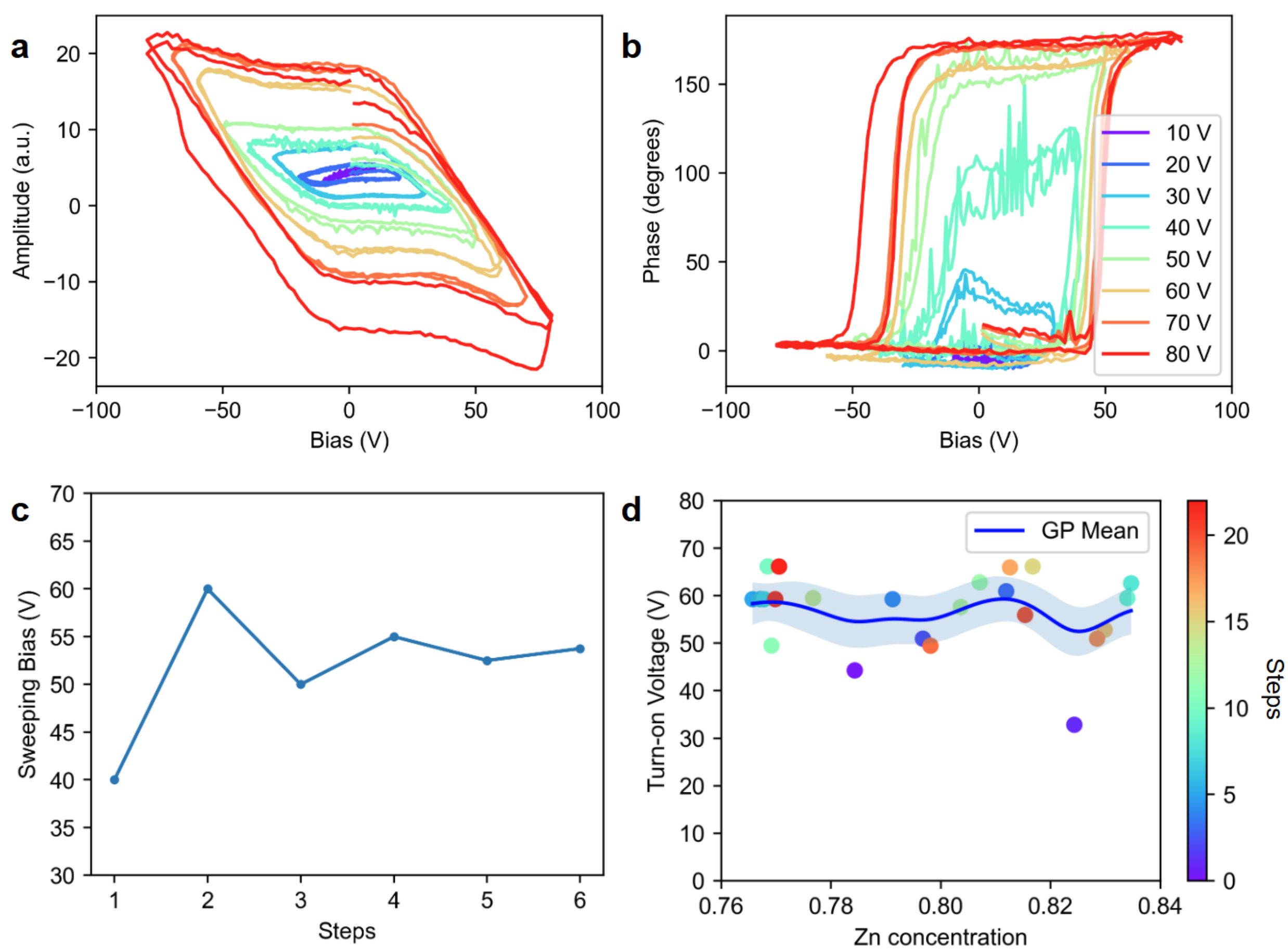

**Fig. S7 | Automated exploration of turn-on voltage in ZMO. a,** DART spectroscopy of piezoresponse amplitude vs. bias curves for sweeping bias ranging from 10 V to 80 V. **b,** phase vs. bias curves for the same set of DART spectra. Full switching of phase occurs between 40 V and 50 V and thus the measured turn-on voltage at this location is also between 40 and 50 V. **c,** Schematic of binary search of turn-on voltage. The search starts with an initial sweeping bias range

of 40 V, and voltage step size of 20 V. At each step, the voltage step size is reduced by a factor of 2 and the direction of voltage change is determined by whether there is 180-degree phase flip in the phase curve (+ for no flip and – for flip). In this example, after 6 search steps, the turn-on voltage is found to be 53.75 V +/- 0.625 V. **d,** vBO exploration of the turn-on voltage with 6-step binary search voltage as the rewards. Here we took 5 seeding points and 25 exploration steps. The UCB acquisition function with beta=100 is used to determine where to measure next at each step. We intentionally excluded the exploration points with turn-on voltage above 75 V as it indicates that either the sample is not switchable or requires too-high voltage to switch, which is dangerous to the probe.

## IV. Phase diagram of (Al,Sc,B)N library and orchestration of SPM with Photoluminescence

To visualize the measured ferroelectric properties as a function of compositions in a standard ternary phase diagram plot, we measured the composition across the library with wavelength dispersive spectroscopy (WDS). The measured B (0.3 – 3.1 %) and Sc (10 – 23.4 %) (colored dots in Fig. S8a-b) concentrations match our estimation of the composition range according to previous X-ray photoelectron spectroscopy (XPS) calibration of our sputtering system on similar samples: B of 0-4 % and Sc of 0-30 %. The concentration of Al was derived from the concentration of B and Sc as c(Al) = 1 – c(B) – c(Sc). To plot the full phase diagram of the combinatorial library, we first interpolate the concentrations of B and Sc across the whole wafer. We have used a linear radial basis function (RBF) with a small smoothing factor of 0.1 to ensure a smooth 2D fit of the compositional distribution. The interpolated distributions of B and Sc are shown in Fig. S8a-b, from which the concentration of Al can be obtained. Then the PFM measured properties can be plotted directly as a function of compositions as in Fig. 5g-i.

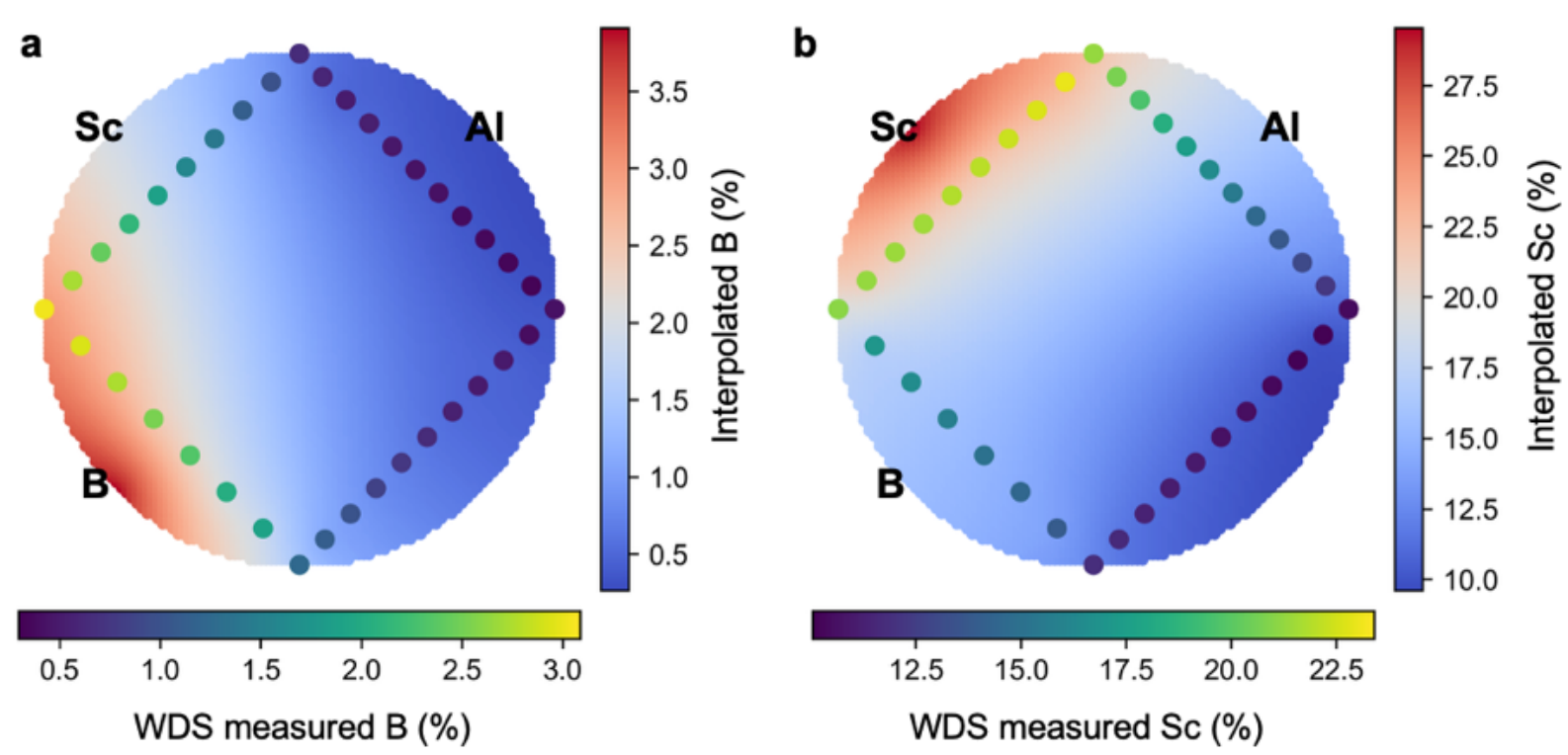

**Fig. S8 | Compositional characterization of (Al,Sc,B)N library with WDS. a-b,** WDS measured concentration of **(a)** B and **(b)** Sc. The distribution of concentrations across the whole sample (colored maps) are interpolated based on the WDS measurements (colored dots).

Furthermore, the photoluminescence (PL) response of the ternary (Al,Sc,B)N sample was mapped across the compositional gradients to develop insights on how the film quality probed by topography images informs the piezoresponse. The PL spectra collected across the thin film library displays two primary features: a low energy peak centered around ~2.7 eV and a higher energy peak centered around ~3.8 eV, as shown in Fig. S9b-j, with center energies shifting with composition-induced bandgap variations across the sample. Emission features at such energies are typically interpreted as related to deep-level defect states involving isolated aluminum vacancies ($V_{Al}$) or complexes formed by aluminum vacancies coupled with oxygen impurities ($V_{Al}$-$O_N$)[60-63]. A similar origin is presumed in interpreting the spectra obtained under UV excitation. Since such deep level states carry a common contributor in the form of Al vacancies (or more generally, cation vacancies), the intensity of the PL spectra is integrated from the 2 eV and 4 eV range with the net integrated intensity across the wafer shown in Fig. S9a.

Comparison between the Al vacancy concentration shown in Fig. 5j with MOBO predicted piezoresponse based on automated SPM measurements in Fig. 5e reveals that at small sample roughness regions which corresponding to large piezoresponse, there is a weaker photoluminescence signal associated with vacancies on the cation site. Therefore, our observed negative correlation between sample roughness and piezoresponse in Fig. 5a-c can be explained by the following pathway: the film height variation is an indication of film quality, which controls the density of lattice defects as indicated by the PL measurement, and thus controls the piezoresponse behavior.

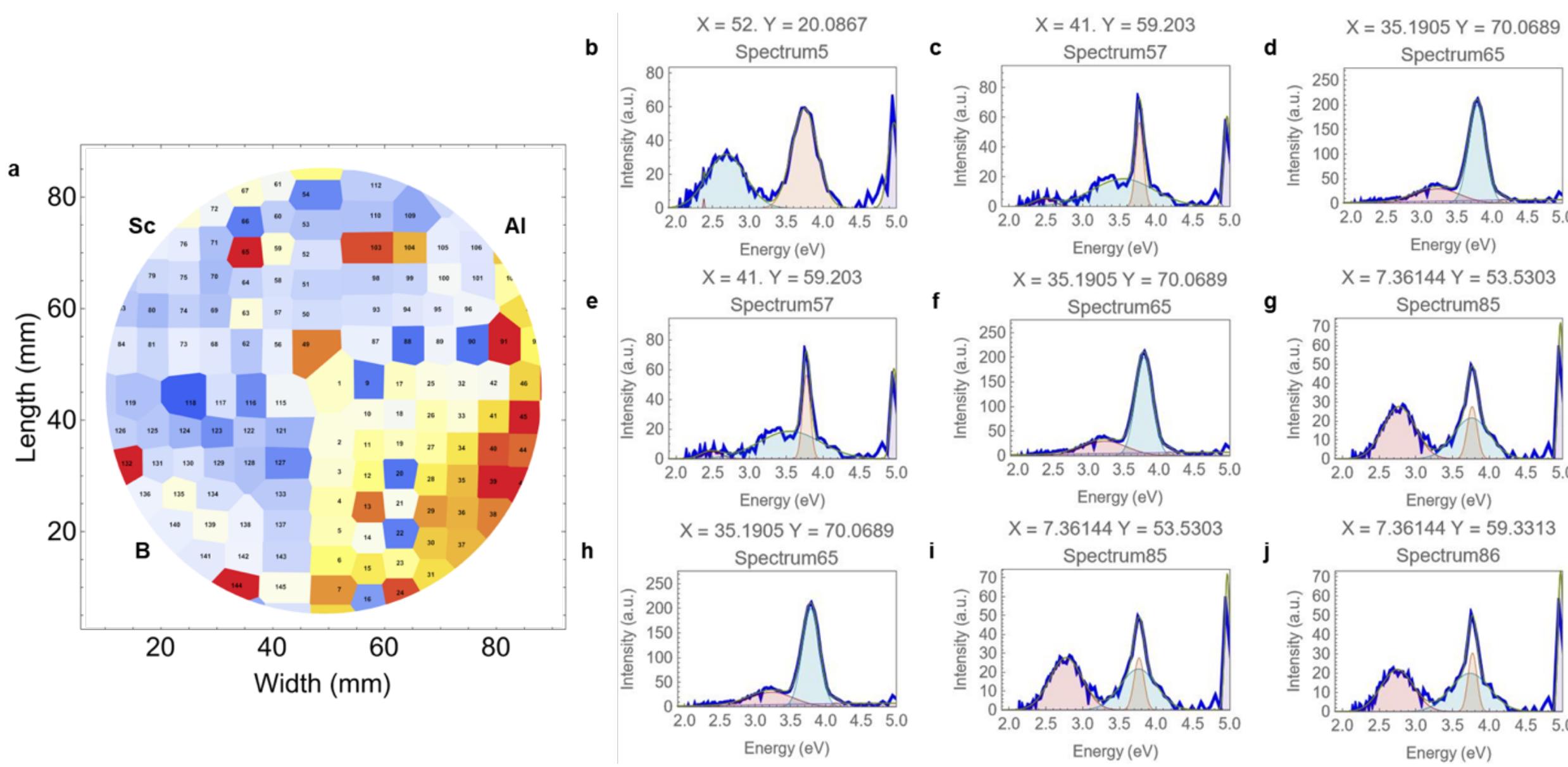

**Fig. S9 | Correlation between Photoluminescence and SPM. a,** the Al vacancy concentration map extracted by integrating the three peaks between 2 eV and 4 eV in the PL spectra acquired across the whole sample. Each numbered pad represents one PL measurement spot. **b-j,** nine representative PL intensity spectra measured at different locations on the (Al,Sc,B)N sample. The three PL peaks between 2 eV and 4 eV are all associated with Al vacancy states. The colored dashed areas show the Gaussian fitting of peaks.